\pdfoutput=1 

\documentclass[preprint,review,3p,times,sort&compress,12pt]{elsarticle}

\usepackage{amssymb}
\usepackage{hyperref}
\usepackage{breakurl,epstopdf,xcolor}

\newenvironment{dedication}
  {
   \thispagestyle{empty}
   \itshape             
   \raggedleft          
  }

\journal{}

	\usepackage{fancyheadings}
\renewcommand{\thispagestyle}[1]{} 

\begin{document}

\begin{frontmatter}



\title{Hubbard pair cluster with elastic interactions. Studies of thermal expansion, magnetostriction and electrostriction}

\begin{dedication}
In Memory of Professor Leszek Wojtczak
\end{dedication}


\author{T. Balcerzak}
\ead{tadeusz.balcerzak@uni.lodz.pl}
\author{K. Sza{\l}owski\corref{cor1}}
\ead{karol.szalowski@uni.lodz.pl}
\address{Department of Solid State Physics, Faculty of Physics and Applied Informatics,\\
University of \L\'{o}d\'{z}, ul. Pomorska 149/153, 90-236 \L\'{o}d\'{z}, Poland}

\cortext[cor1]{Corresponding author}

\date{\today}

\begin{abstract}

The pair cluster (dimer) is studied within the framework of the extended Hubbard model and the grand canonical ensemble. The elastic interatomic interactions and thermal vibrational energy of the atoms are taken into account.
The total grand potential is constructed, from which the equation of state is derived. In equilibrium state, the deformation of cluster size, as well as its derivatives, are studied as a function of the temperature and the external magnetic and electric fields. In particular, the thermal expansion, magnetostriction and electrostriction effects are examined for arbitrary temperature, in a wide range of Hamiltonian parameters. 

\end{abstract}


\end{frontmatter}

\section{Introduction}

The Hubbard model~\cite{Anderson1959,Hubbard1963,Gutzwiller1963,Kanamori1963}
plays an important role in contemporary solid state physics. Since its
formulation, numerous applications of the model have been developed and the
model itself has been extensively investigated as a fundamental, prototypic
model for description of correlated
fermions~\cite{Nolting2009,Yosida1998,Tasaki1998,Mielke2015}. Many of the
studies were concentrated on the typical three-dimensional, two-dimensional or
one-dimensional Hubbard model (e.g.~\cite{Shastry1986,Hirsch1989,Fuchs2011}).
However, with an increase of interest in studies of nanoclusters, which is
stimulated by the development of nanotechnology, intensive investigations of
the Hubbard model applied to the systems containing small number of atoms have been
performed. Some of them involve the studies of small clusters with either exact
or close to exact
methods~\cite{Ohta1994,Ojeda-Lopez1997,Noce1997,Ojeda1999,Becca2000,Ricardo-Chavez2001,Hancock2002,Hancock2005,Schumann2007,Schumann2008,Yang2009,Schumann2010,Feldner2010,Ovchinnikova2013,Hancock2014,Souza2016,Szalowski2017,Szalowski2018a}.
Among such systems, a two-atomic cluster (dimer) plays a special role, being
the smallest nanosystem where the Hubbard model can be adopted, and the exact
analytical solutions can be found. As a consequence, numerous aspects of
physics of Hubbard dimers focused considerable attention in the
literature~\cite{Alvarez-Fernandez2002,Harris1967,Wortis2017,Cheng1976,Longhi2011,Juliano2016,Kozlov1996,Silantev2015,Hasegawa2005,Hasegawa2011,Spalek1979,Balcerzak2017,Balcerzak2018,Balcerzak2018a,Ullrich2018a,Sagredo2018,Carrascal2015,Fuks2014,Kamil2016,Balasubramanian2017}.
These facts serve as a sound motivation for further comprehensive study of the
two-atomic Hubbard system aimed at its full and exact thermodynamic
description. In line with this trend, recently the Hubbard pair cluster
embedded simultaneously in the external magnetic and electric fields has been
studied by the exact analytical methods in the framework of the grand canonical
ensemble. In this approach the system is open, so that it is able to exchange
the electrons with its
environment~\cite{Balcerzak2017,Balcerzak2018,Balcerzak2018a}. The statistical
and thermodynamic properties, both magnetic and electric ones, have been
extensively studied there. 

Despite these investigations, there is still room for extending the model of
interest. Supplementing purely electronic models by including further degrees
of freedom can cause new kinds of behaviour to emerge and new control
parameters can be gained this way. Purely electronic models can be
supplemented, for example, with subsystems composed of localized spins
(e.g.~\cite{Cisarova2014,Nalbandyan2014,Galisova2015,Galisova2015a,Cencarikova2016,Ananikian2017,Cencarikova2018,Sousa2018}),
which are still exactly solvable. Immersing of such system in the external
fields can give access to unique properties (like, for example, manifestations
of the magnetoelectric effect~\cite{Cencarikova2018}). Enriching the purely
electronic models is also a step towards more complete characterization of the
real physical systems. A good example can be inclusion of the Hamiltonian
terms, which are responsible for coupling of the electronic degrees of freedom
with the lattice. Such a study, aimed at the description of magnetomechanical
and electromechanical properties of the Hubbard dimer, has not been undertaken
yet. Therefore, the aim of the present paper is extension of the recent
studies, basing on the formalism developed in Ref.~\cite{Balcerzak2017}, by
including elastic interactions in the description of the Hubbard dimer. Such
interactions are especially important since they are responsible for stability
of the dimer structure and allow to account for the energy of thermal
vibrations. It can be expected that after combining with the Hubbard
Hamiltonian, the elastic interactions will lead to new phenomena, connecting
the magnetic, electric and mechanical properties. 

The elastic interaction between the atoms is assumed here in the form of the
Morse potential~\cite{Morse1929,Girifalco1959}. The thermal vibrational energy
is obtained in the quasi-harmonic approximation, where the anharmonicity is
taken into account by the Gr\"{u}neisen
parameter~\cite{Gruneisen1912,Balcerzak2010}. In turn, the Hubbard pair
Hamiltonian is taken in its extended form, with the Coulomb interaction between the electrons on different atoms and the hopping integral depending on dimer size.

On the basis of the above assumptions, the total grand thermodynamic potential is constructed, from which all statistical and thermodynamic properties can be obtained in a self-consistent manner. In particular, the deformation of interatomic distance (dimer length), magnetostriction and electrostriction coefficients are calculated for various temperatures, and the effect of the external magnetic and electric fields on the mentioned quantities is also investigated. In addition, the chemical potential is studied in the presence of elastic interactions, showing a new behaviour in comparison with the conventional Hubbard model. A special attention is paid to the low-temperature region, where the discontinuous quantum changes of these properties can be demonstrated.

The paper is organized as follows:  In the next Section the theoretical model is presented and the basic formulae, important for numerical calculations, are derived. In the successive Section the selected results of calculations are illustrated in figures and discussed. 
The last Section is devoted to a brief summary of the results and final conclusions.

 \section{Theoretical model}

In the present section, a step-by-step development of the theoretical model for the Hubbard dimer including the elastic and vibrational properties is presented and the solution of the model is described. 

\subsection{Grand potential for the Hubbard dimer}

Extended Hubbard Hamiltonian for the pair of atoms $(a,b)$ embedded in the external fields is of the form:
\begin{eqnarray}
\mathcal{H}_{a,b}&=&-t\sum_{\sigma=\uparrow,\downarrow}\left( c_{a,\sigma}^+c_{b,\sigma}+c_{b,\sigma}^+c_{a,\sigma} \right)+U\left(n_{a,\uparrow}n_{a,\downarrow}+n_{b,\uparrow}n_{b,\downarrow}\right)\nonumber\\
&& -H\left(S_a^z+S_b^z\right) -V\left(n_{a}-n_{b}\right) + W n_{a}n_{b},
\label{eq1}
\end{eqnarray}
where $t>0$ is the hopping integral, $U\ge0$ is on-site Coulomb repulsion and $W$ stands for the Coulomb interaction between electrons localized on $a$ and $b$ atoms. In \ref{eq1},  $H=-g\mu_{\rm B}H^z$ denotes the magnitude of an external uniform magnetic field $H^z$, whereas the electric field $E$ is determined by the electrostatic potentials $V_{a}=-V_{b}=V$ applied to both atoms. Thus, $E=2 V/\left(| e| d\right)$, where $d$ is the interatomic distance and $e$ is the electron charge. The quantity $W$ in \ref{eq1} is assumed in the classical form: $W=1/(4 \pi \epsilon_0)\, e^2/d$.

The creation ($c_{\gamma,\sigma}^+$) and annihilation ($c_{\gamma,\sigma}$) operators for site $\gamma = a,b$ and spin state $\sigma$ can be used to construct the occupation number operators
$n_{\gamma,\sigma}$ for the electrons with given spin:
\begin{equation}
n_{\gamma,\sigma}=c_{\gamma,\sigma}^{+}c_{\gamma,\sigma}.
\label{eq2}
\end{equation}
The total occupation number operators at the site $\gamma = a,b$ are then given by:
\begin{equation}
n_{\gamma}=n_{\gamma,\uparrow}+n_{\gamma,\downarrow}.
\label{eq3}
\end{equation}
Moreover, the spin operators $S_{\gamma}^z$ are defined as follows:
\begin{equation}
S_{\gamma}^z=\left(n_{\gamma,\uparrow}-n_{\gamma,\downarrow}\right)/2.
\label{eq4}
\end{equation}
Since we intend to study the elastic properties of the model, the interatomic distance $d$ can be presented as follows:
\begin{equation}
d=d_{0}\left(1+\varepsilon \right),
\label{eq5}
\end{equation}
where $d_{0}$ is the equilibrium distance defined in the absence of the external fields ($H=0, \, E=0$) and for zero temperature ($T=0$). In \ref{eq5}, $\varepsilon$ defines the temperature and fields-induced deformation. As it can be noted, some of the coefficients in Hamiltonian \ref{eq1}, like $V$, $W$ and $t$, are interatomic distance-dependent. For instance, for the electric potential we have:
\begin{equation}
V=V_{0}\left(1+\varepsilon \right),
\label{eq6}
\end{equation}
where $V_{0}=E| e| d_{0}/2$. In turn, for the interatomic Coulomb interaction we can write:
\begin{equation}
W=\frac{W_{0}}{1+\varepsilon},
\label{eq7}
\end{equation}
where $W_{0}=\frac{e^{2}}{4 \pi \epsilon_0 d_{0}}$. Moreover, for the hopping integral we assume the power-law distance dependence, namely:
\begin{equation}
t=t_{0}\left(\frac{d}{d_{0}}\right)^{-n}= t_{0}\left(1+\varepsilon \right)^{-n},
\label{eq8}
\end{equation}
with a constant power index $n$, which can be used if the deformation $\varepsilon$ is small.

We treat the Hubbard pair as an open system, with a variable number of electrons, which can be exchanged with the environment. It corresponds to the physical situation when, for instance, the cluster is interacting with the metallic substrate, which forms the electronic reservoir, and the average number of electrons in the system results from the thermodynamic equilibrium. In such small system the relative fluctuations of the number of electrons, when related to its statistical mean value, can be significant. Therefore, in this case a formalism of the grand canonical ensemble should be used, which is more appropriate here than the canonical ensemble, with a fixed number of electrons.

 In the framework of the grand canonical ensemble, the Hamiltonian \ref{eq1}
should be extended by adding the term $-\mu \left(n_a
+n_b\right)$, where $\mu$ is the
chemical potential. For such extended Hamiltonian, the diagonalization
procedure can be performed exactly using the method outlined in
Ref.~\cite{Balcerzak2017}. It is worth noticing that by including the new term,
$W n_{a}n_{b}$, the complexity of diagonalization procedure will not increase in
comparison to that presented in Ref.~\cite{Balcerzak2017}. We note that
$n_{a}n_{b}$ is a diagonal {16~$\times$~16} matrix and corresponds to 16 possible
states.
Its diagonal elements are: $a_{1,1}=a_{2,2}=a_{3,3}=a_{4,4}=a_{5,5}=0$, $a_{6,6}=a_{7,7}=1$, 
$a_{8,8}=2$, $a_{9,9}=0$, $a_{10,10}=a_{11,11}=1$, $a_{12,12}=2$, $a_{13,13}=0$, 
$a_{14,14}=a_{15,15}=2$, $a_{16,16}=4$.
Such a matrix, multiplied by $W/t$, should be simply added to that
presented in Ref.~\cite{Balcerzak2017} (Eq. B.5 therein).

 As a result of diagonalization, the grand thermodynamic potential $\Omega_{a,b}$ for the Hubbard pair  can be found from the general formula:
\begin{equation}
\Omega_{a,b}=-k_{\rm B}T \ln \{{\rm Tr}_{a,b} \,\exp \lbrack -\beta \left(\mathcal{H}_{a,b}-\mu\left(n_a+n_b\right)\right)\rbrack \}.
\label{eq9}
\end{equation}
If we denote the mean number of the electrons per atom by $x$, then the following relation is satisfied:
\begin{equation}
2x=\left\langle n_a\right\rangle +\left\langle n_b\right\rangle =-\left(\frac{\partial \Omega_{a,b}}{\partial \mu}\right)_{T,H,E}.
\label{eq10}
\end{equation}
In Eq.~\ref{eq10}, $\left\langle \cdots\right\rangle $ means the statistical average, which is carried out over all 16 states, and is dependent on the temperature and external fields.
From Eq.~\ref{eq10} the chemical potential $\mu$ can be found, provided that $x$ is established, and then the expression \ref{eq9} is complete.

\subsection{Elastic (static) interaction}

Irrespective of the Hubbard contribution, the interatomic static interaction of
the core electrons can be taken into account in the form of the Morse potential
energy~\cite {Morse1929}:
\begin{equation}
V_{M}\left(r\right)=D\left[1-e^{-\alpha\left(r-r_0\right)/r_0}\right]^{2}.
\label{eq11}
\end{equation}
In Eq.~\ref{eq11}, $D$ is the potential depth, parameter $\alpha$ controls the potential width and its asymmetry, whereas $r_{0}$ is the equilibrium distance (potential minimum position) for this interaction at $T=0$. The static energy can be written in terms of deformation $\varepsilon$, and its value can be normalized by the requirement that $V_{M}\left(\varepsilon =0\right)=0$. In this way we obtain the elastic energy in the form of:
\begin{equation}
V_{M}\left(\varepsilon \right)=D\left[1-e^{-\alpha \left(\frac{d_0}{r_0}\left(1+\varepsilon \right)-1 \right)}\right]^{2} - D\left[1-e^{-\alpha \left(\frac{d_0}{r_0}-1 \right)}\right]^{2}.
\label{eq12}
\end{equation}
The dimensionless parameter $d_0/r_0$ can be found from the minimum condition of the total energy, taken at $T=0$ in the absence of the external fields.

\subsection{Vibrational (thermal) energy}

In order to take into account the thermal vibrations, the two-atomic system can
be treated as a quantum oscillator having reduced mass and described in the
centre of mass coordinate system. This situation corresponds, for instance, to
interaction of the system with the solid substrate, which forms a phononic
bath. For the sake of simplicity we adopt the quasi-harmonic Einstein model, in
which the anharmonicity is taken into account by the Gr\"uneisen parameter
$\Gamma$~\cite{Balcerzak2010}. The vibrational free energy of the Einstein
oscillator is then given by:
\begin{equation} 
E_{\omega} =k_{\mathrm B} T \ln \left[2 \sinh \left(\frac{1}{2}\beta \hbar \omega \right) \right].
\label{eq13}
\end{equation}
The frequency of oscillations, $\omega$, is dependent on the interatomic distance and can be expressed as a function of the deformation $\varepsilon$ as follows:
\begin{equation} 
\omega =\frac{\omega_0}{\left(1+\varepsilon \right)^{\Gamma}},
\label{eq14}
\end{equation}
where the Gr\"uneisen parameter $\Gamma$ satisfies the relationship:
\begin{equation} 
\Gamma =-\frac{d}{\omega}\left(\frac{\partial \omega}{\partial d}\right).
\label{eq15}
\end{equation}
It is known that $\Gamma$ can be determined theoretically for several
interatomic potentials and different dimensionalities of the system. For
instance, for the Morse potential and 1D system
$\Gamma=\left(3/2\right) \alpha$~\cite{Krivtsov2011}. This convenient relationship will be adopted
in further calculations.

Thus, the vibrational energy can be finally presented as:
\begin{equation} 
E_{\omega} =k_{\mathrm B} T \ln \left[2 \sinh \left(\frac{t_0}{k_{\mathrm B} T}\frac {\hbar \omega_0}{t_0} \frac{1}{2\left(1+\varepsilon \right)^{\Gamma}} \right) \right],
\label{eq16}
\end{equation}
where $k_{\mathrm B} T/ t_0$ is the dimensionless temperature, and $\hbar \omega_0/ t_0$ is the dimensionless characteristic energy parameter depending on the atomic mass.

\subsection{The equation of state (EOS)}

The total grand potential, $\Omega$, of the system in question can be constructed as a sum of the Hubbard grand potential \ref{eq9}, elastic (static) energy \ref{eq12} and vibrational Helmholtz free-energy \ref{eq16}:
\begin{equation} 
\Omega=\Omega_{a,b}+V_{M}\left(\varepsilon \right)+E_{\omega}. 
\label{eq17}
\end{equation}
This grand potential should be minimized first with respect to the deformation $\varepsilon$ by imposing the necessary equilibrium condition:
\begin{equation} 
\left(\frac{\partial \Omega}{\partial \varepsilon}\right)_{T,H,E}=0. 
\label{eq18}
\end{equation}
Differentiation of $\Omega$ will then lead to three forces arising out of the three terms in r.h.s. of Eq.~\ref{eq17}.

The Hubbard force, $F_{a,b}$, is connected with the change of the Hubbard grand potential under the  deformation:
\begin{equation} 
F_{a,b}=-\frac{1}{d_0}\left(\frac{\partial \Omega_{a,b}}{\partial \varepsilon}\right). 
\label{eq19}
\end{equation}
With the help of Eq.~\ref{eq9} we get:
\begin{equation} 
-d_0F_{a,b}=\frac{\partial }{\partial \varepsilon}\left\langle \mathcal{H}_{a,b}\right\rangle -2x\frac{\partial \mu}{\partial \varepsilon}, 
\label{eq20}
\end{equation}
where $\left\langle \mathcal{H}_{a,b}\right\rangle $ is the statistical average of the Hubbard Hamiltonian. The energy 
$\left\langle \mathcal{H}_{a,b}\right\rangle $ can be found exactly after the process of diagonalization. Then, the derivative $\partial \left\langle \mathcal{H}_{a,b}\right\rangle / \partial \varepsilon$ can be calculated analytically with the help of Eqs. \ref{eq6}--\ref{eq8}. In turn, the derivative $\partial \mu/\partial \varepsilon$ can be, in principle, calculated numerically for the arbitrary concentration $\left(0\le x\le 2\right)$ of Hubbard electrons. For such purpose Eq.~\ref{eq10} for the chemical potential can be used. 
Interestingly, for $x=1$, i.e.,~for half-filled energy levels, the analytical formula for the chemical potential $\mu$ can be found, namely:
\begin{equation} 
\mu=\frac{W_0}{1+\varepsilon}+\frac{U}{2}, 
\label{eq21}
\end{equation}
which effectively facilitates the calculations of $\partial \mu/\partial \varepsilon$. As a result, the Hubbard force for $x=1$ can be presented in the following analytic form:
\begin{eqnarray} 
F_{a,b}&=&-\frac{n}{\left(1+ \varepsilon \right)^{n+1}}\frac{t_0}{d_0}\sum_{\sigma=\uparrow,\downarrow}\left( \left\langle c_{a,\sigma}^+c_{b,\sigma}\right\rangle +\left\langle c_{b,\sigma}^+c_{a,\sigma}\right\rangle  \right)\nonumber\\
&&+ \frac{V_0}{d_0}\left(\left\langle n_{a}\right\rangle -\left\langle n_{b}\right\rangle \right)-
\frac{W_0}{d_0\left(1+\varepsilon \right)^2}\left( \left\langle n_{a}\right\rangle +\left\langle n_{b}\right\rangle  
-\left\langle n_{a}n_{b}\right\rangle \right).
\label{eq22}
\end{eqnarray}
The statistical averages in Eq.~\ref{eq22} can be calculated after
performing the diagonalization of the Hamiltonian. The examples of similar
calculations (in the absence of elastic interactions) have already been
presented in Ref.~\cite{Balcerzak2018}.

As a next step, the elastic force, $F_{\varepsilon}$, is found from the formula:
\begin{equation} 
F_{\varepsilon}=-\frac{1}{d_0}\left(\frac{\partial V_{M}\left(\varepsilon \right)}{\partial \varepsilon}\right), 
\label{eq23}
\end{equation}
where $V_{M}\left(\varepsilon \right)$ is given by Eq.~\ref{eq12}. Hence, we obtain:
\begin{equation} 
F_{\varepsilon}=-\frac{2 D \alpha}{r_0}\left[1-e^{-\alpha \left(\frac{d_0}{r_0}\left(1+\varepsilon \right)-1 \right)}\right] e^{-\alpha \left(\frac{d_0}{r_0}\left(1+\varepsilon \right)-1 \right)}. 
\label{eq24}
\end{equation}

The last force, $F_{\omega}$, which is of vibrational origin, is defined as:
\begin{equation}
F_{\omega}=-\frac{1}{d_0}\left(\frac{\partial E_{\omega}}{\partial \varepsilon}\right), 
\label{eq25}
\end{equation}
where $E_{\omega}$ is given by Eq.~\ref{eq16}. From Eqs. \ref[nosort]{eq25,eq16} we obtain:
\begin{equation}
F_{\omega}=\frac{\Gamma}{2\left(1+ \varepsilon \right)^{\Gamma+1}}\frac{\hbar \omega_0}{d_0}
\tanh^{-1}\left(\frac{t_0}{k_{\mathrm B} T}\frac {\hbar \omega_0}{t_0} \frac{1}{2\left(1+\varepsilon \right)^{\Gamma}} \right). 
\label{eq26}
\end{equation}
It follows from the numerical calculations that the Hubbard force, $F_{a,b}$, is compressive (negative) around $\varepsilon =0$. On the other hand, the elastic force, $F_{\varepsilon}$, is compressive for $d>r_0$ and expansive (positive) for $d<r_0$, whereas the vibrational force, $F_{\omega}$, is always of expansive type. The necessary condition for thermodynamic equilibrium requires the balance of all these forces, namely:
\begin{equation}
F_{a,b} + F_{\varepsilon} + F_{\omega}=0. 
\label{eq27}
\end{equation}
Expression \ref{eq27} is the EOS for the Hubbard dimer with the elastic interactions and thermal vibrations taken into account. It enables calculation of spontaneous deformation, $\varepsilon$, of the dimer length for given values of the intensive parameters: $T$, $H$ and $E$, whereas the system is in stable equilibrium. At first, in the absence of deformation (i.e.,~when $\varepsilon =0$ for $T=0$, $H=0$ and $E=0$), the constant parameter $d_0/r_0$ should be found from Eq.~\ref{eq27}, giving the equilibrium distance $d_0$ at the ground state.

The numerical calculations based on the EOS \ref{eq27} will be presented in the next Section.

 \section{Numerical results and discussion}

The numerical results will be presented for the half-filling case, i.e.,~when
the average number of electrons per atom is $x=1$. In the Hubbard
Hamiltonian we select the exponent $n$ in the hopping integral (\ref{eq8}) equal to $n=6$ in order to describe a rapid variability of the
function around $d_0$. It is in analogy with the possible dependence
of the exchange integral vs. the distance~\cite{Balcerzak2017a} and refers to
the attractive part of the Lennard-Jones potential. A more accurate description
of the hopping integrals dependence on the bond length can be given by the
Goodwin, Skinner and Pettifor function~\cite{Goodwin1989}. The interatomic
Coulomb interaction parameter $\frac{W_0d_0}{r_0 t_0}=\frac{e^{2}}{4 \pi \epsilon_0 r_{0}t_0}$ can be estimated for the realistic
values of $r_0 \approx 1.5 ~\AA$ and $t_0 \approx 2.5~{\rm eV}$ giving a figure of $\approx$ 3.83,
and for further calculations a round value of $4$ is accepted.
All dimensionless energies, temperature and fields magnitudes are normalized by
$t_0$ and the interatomic equilibrium distance $d_0$ is related
to the $r_0$ constant. In this convention, we chose the value
$D/t_0=4$ for the Morse potential depth, while the asymmetry parameter is
assumed with its typical value $\alpha=4$~\cite{Girifalco1959}. For the
thermal vibrations we assume the characteristic energy parameter $\hbar \omega_0/ t_0=0.5$,
whereas the Gr\"uneisen parameter is $\Gamma=\left(3/2\right) \alpha$~\cite{Krivtsov2011}. Such a
set of parameter values, although not describing any specific material, seems
physically acceptable and can be used to demonstrate how the formalism works
when the numerical calculations are performed. 
We are aware that the choice of these values can be discussed and modified
according to the physical situation. Unfortunately, because of the limited
scope of the paper we are not able to present the influence of all these
parameters on the numerical results. However, it can be mentioned that, for
instance, some change of the asymmetry parameter $\alpha$, or the potential
depth $D$, may have remarkable influence on the thermodynamic
properties, as it has been demonstrated in Ref. \cite{Balcerzak2018b}.

For the above set of parameters and for $U/t_0=0$, from Eq.~\ref{eq27} we obtain the equilibrium distance $d_0/r_0=0.922085$ in the ground state (i.e.,~for $T=0$, $H=0$ and $E=0$, whereas deformation is $\varepsilon =0$). Analogously, for $U/t_0=1,\,2,\,5$ and $10$, the corresponding equilibrium distances are: $d_0/r_0=0.918756$, 0.915598, 0.915319 and 0.941767, respectively. At the same time, in the ground state the chemical potentials, $\mu/t_0$, for $U/t_0=0,\,1,\,2,\,5$ and $10$ are as follows: $\mu/t_0=4.337994$, 4.853712, 5.368729, 6.870059 and 9.247335, respectively.

We see from the above results that the equilibrium interatomic distance
$d_0$ is not a monotonous function of the parameter $U$.
Moreover, the chemical potential $\mu$ is not equal to $U/2$, as
it happens in the case when the interatomic Coulomb repulsion, as well as the
elastic interactions, are neglected~\cite{Nolting2009,Balcerzak2017}. The
values of $\mu$ obtained here are in agreement with Eq.~\ref{eq21}.
The formula \ref{eq21} also predicts that out of the ground state, i.e.,~for
$T>0$,  and in the possible presence of the external fields, the
chemical potential is no longer constant, but it should be a function of
$T$, $H$ and $E$, via its dependency on the
deformation $\varepsilon$. This new behaviour is illustrated in Fig.~\ref{fig1}-\ref{fig3}. For instance, in \ref{fig1} the difference $(\mu-U/2)/t_0$ is shown
as a function of the dimensionless temperature $k_{\rm B}T/t_0$ for several values
of $U/t_0$, when the external fields are absent. In general, the functions
decrease with the increasing temperature and, at the same time, the dependency of
$(\mu-U/2)$ on $U$ is not monotonous. In \ref{fig2} also the
decreasing tendency of $(\mu-U/2)/t_0$ as a function of dimensionless magnetic
field $H/t_0$ is shown for the same values of $U/t_0$. In this case
the electric field is  equal to $E=0$, and the moderate temperature,
$k_{\rm B}T/t_0=0.5$, is assumed. In \ref{fig3} the chemical potential is plotted for
$k_{\rm B}T/t_0=0.5$ and $H=0$, as a function of the dimensionless electric
field $E| e| r_{0}/t_0$. Apart from the decreasing tendency of $(\mu-U/2)/t_0$ vs.
$E$, which is visible for small $U$, the increasing tendency
of the chemical potential for $U/t_0=8$ and $U/t_0=10$ is worth noticing.

\ref{fig4} presents relative deformation $\varepsilon$ of the dimer size vs.
magnetic field $H/t_0$ at very low temperature, $k_{\rm B}T/t_0=0.001$, whereas
$E=0$. Several values of $U/t_0$ are examined. As we know from our
previous studies~\cite{Balcerzak2017,Balcerzak2018,Balcerzak2018a}, in the
low-temperature region the quantum phase transitions should occur at the
magnetic (or electric) critical fields. Such transitions are of the first
order, as regards the behaviour of total dimer magnetization, which results
from the change from the singlet state, where $\left\langle  S^z_a + S^z_b \right\rangle  =0$, to a triplet one,
where $\left\langle  S^z_a + S^z_b \right\rangle  =1$. The rapid change of magnetization is accompanied by
instantaneous changes of the remaining thermodynamic quantities. In \ref{fig4}
the changes of deformation $\varepsilon$ are illustrated at the transition from
the singlet state, where $\varepsilon=0$, to the triplet (ferromagnetically
saturated) state. It should be mentioned that due to the presence of
interatomic Coulomb interaction $W$, the critical fields $H_c$
found here have different values in comparison with those from our previous
papers~\cite{Balcerzak2017,Balcerzak2018,Balcerzak2018a}. Unfortunately, there
is no analytical formula for these critical fields, and they can be calculated
numerically only by equating the total grand potentials of neighbouring phases.

In \ref{fig5} the electrostriction coefficient, $\nu_{T,H}=\frac{1}{d}\left(\frac{\partial d}{\partial E}\right)_{T,H}=\frac{1}{1+\varepsilon}\left(\frac{\partial \varepsilon}{\partial E}\right)_{T,H}$, is presented in
dimensionless units, $\frac{\nu_{T,H} t_0}{| e| r_0}$, as a function of the magnetic field
$H/t_0$. Several values of the parameter $U$ are examined at the
temperature equal to $k_{\rm B}T/t_0=0.001$, whereas the electric field is constant and equal to $E | e| r_0/t_0=0.5$. As it is seen, the characteristic steps corresponding to the
quantum phase transitions occur for the electrostriction coefficient. The
critical magnetic fields $H_c$, where the jumps occur, decrease
with an increase in the Coulombic repulsion energy $U$, which effect is
analogous to that noted in the previous figure. Moreover, for higher
$U$ values, the existence of negative electrostriction is
demonstrated. It can also be noted that the electrostriction coefficient takes
zero value in the triplet (ferromagnetically saturated) state, because then the
electric field is not able to change the charge
distribution~\cite{Balcerzak2018a}, whereas the electric potential is
compensated by the chemical potential. As a result, the total energy is not
influenced by the electric field and length of the dimer remains unchanged.

In the next figure, \ref{fig6}, the deformation $\varepsilon$ is presented at
the temperature $k_{\rm B}T/t_0=0.001$ and the magnetic field $H/t_0=4.8$ as a function
of dimensionless electric field $E | e| r_0/t_0$. Again, the characteristic jumps
of $\varepsilon$ are shown for various curves corresponding to different values
of $U$. The critical electric fields $E_c$, where the phase
transition from triplet to singlet state takes place, increase with an
increase in $U$. Below the critical electric field, i.e.,~in triplet
state, the deformation $\varepsilon$ is constant vs. $E$, which results
in null electrostriction effect. This fact supports the discussion presented in
the preceding paragraph and is in agreement with the general behaviour of the
critical fields for different $U$ calculated for the ground state in
the absence of the elastic interactions~\cite{Balcerzak2018} (Fig. 1 therein).
On the basis of Fig.~\ref{fig3}-\ref{fig6}, it may be interesting to
notice that instead of speaking about the critical fields $E_c$ (or
$H_c$)
for given $U$-parameter and the fields $H$ (or $E$), we can equally well speak about the critical Coulomb potential $U_c$ for given $H$ and $E$.

\ref{fig7} is devoted to illustration of the magnetostriction behaviour in two
temperatures: $k_{\rm B}T/t_0=0.05$ and $k_{\rm B}T/t_0=0.1$ for the same magnetic field as in
previous figure, $H/t_0=4.8$. The magnetostriction coefficient is defined as
$\lambda_{T,E}=\frac{1}{d}\left(\frac{\partial d}{\partial H}\right)_{T,E}=\frac{1}{1+\varepsilon}\left(\frac{\partial \varepsilon}{\partial H}\right)_{T,E}$ and is presented in dimensionless units, $\lambda_{T,E} t_0$, vs.
dimensionless electric field $E | e| r_0/t_0$. It is seen that the magnetostriction
coefficient in lower temperature shows the sharp peaks, whereas in higher
temperature the maxima are lower and much more diffused. These maxima appear
approximately at the critical electric fields $E_c$ and are placed at
different positions for the curves plotted for various parameters $U$.
In particular, when $U$ increases, the critical fields are shifted
towards higher values of $T$. For $T \to 0$ the magnetostriction coefficient is
zero everywhere aside from the singularities, because in the low-temperature
region the magnetic field is not able to change the magnetization, both in
singlet and in triplet state, except for the phase transition
point~\cite{Balcerzak2018}. As a result, the magnetic energy is constant aside
from the singularities and changes of the magnetic field do not influence the
dimer size. This conclusion is also in agreement with \ref{fig4} where, apart from
the phase transition points, the plots of $\varepsilon$ vs. $H$ are
flat, which corresponds to the null magnetostriction effect. When the temperature
increases, the quantum phase transitions become diffused and, as a result, the
magnetostriction coefficient presents broadened maxima with reduced height. The
positions of the peaks are quite stable and are only slightly shifted towards
larger electric fields. At the same time, the height of the peaks strongly
depends on the Hubbard parameter $U$, showing a decrease of maximal
magnetostriction coefficient when $U$ increases. 

The coefficient of the linear thermal expansion is presented in \ref{fig8} as a function of the dimensionless temperature $k_{\rm B}T/t_0$. This coefficient is defined as $\alpha_{H,E} =\frac{1}{1+\varepsilon}\left(\frac{\partial \varepsilon}{\partial T}\right)_{H,E}$, where the derivative over $T$ is taken at constant fields $H$ and $E$. In \ref{fig8} the dimensionless quantity $\alpha_{H,E}\, t_0/k_{\rm B}$ is plotted, whereas the fields $H$ and $E$ are absent. The most interesting effect illustrated in this figure is the maximum of the thermal expansion and its evolution when the Coulombic repulsion energy $U$ increases. This maximum is initially small for the curve corresponding to $U=0$ and gradually disappears for the curves with $U/t_0=1$ and  $U/t_0=2$. Then, for $U/t_0=3$ it reappears at higher temperature and becomes very pronounced for the curves corresponding to $U/t_0=6$ and  $U/t_0=10$. At the same time, the maximum is shifted towards lower temperatures again. The occurrence of this maximum is not connected with the phase transition, since the system still remains in the singlet state. It rather results from a complicated interplay and competition between different contributions to the total grand potential, which depend on the interatomic distance. 

In order to outline the origin of this maximum, it can be noticed that an increase in $U$-potential brings on the charge separation, and as a result the occupation correlation on different atoms, $\left\langle n_a n_b\right\rangle $, increases. For instance, in \ref{fig8}~ for the fixed temperature $k_{\rm B}T/t_0 =0.5$, and $U/t_0=2$, 6 and 10, the occupation correlations amount to: $\left\langle n_a n_b\right\rangle =0.241$, 0.822 and 0.9751, respectively. Then, in order to compensate the increase in interatomic Coulomb energy, the $W$ coefficient (Eq.~\ref{eq7}) must decrease, which means that deformation $\varepsilon$ takes higher values there. One can also note that, if $d_0/r_0<1$, some increase in $\varepsilon$ helps to lower the Morse potential energy.
The resulting deformation is obtained as the solution of EOS (Eq.
\ref{eq27}). For the above example ($k_{\rm B}T/t_0 =0.5$ and $U/t_0=2$, 6 and
10), the corresponding deformations are: $\varepsilon=0.0167$, 0.0358 and 0.0400,
respectively.  On the other hand, for given $U$, the deformation is a
non-linearly increasing function of temperature. In this case the increase in
$\varepsilon$ is also connected with the energy compensation mechanism, and it
is mainly stimulated by the fact that in the singlet (paramagnetic) phase the
increase in temperature results in a nonlinear increase of the mean hopping
energy (see~\cite{Balcerzak2018} and Fig. 5 therein). Then, the derivative
$\partial \varepsilon / \partial T$ can reveal a non-monotonous behaviour vs. temperature and the
maximum of $\alpha_{H,E}$ appears as a consequence.
The non-linear changes of $\varepsilon$ are more pronounced in the intermediate range of temperatures and for larger $U$-parameters, where the competition between various pressure components in EOS is more dynamic. In our opinion, the revealed maximum of $\alpha_{H,E}$ is, to some extent, analogous to the paramagnetic maximum of the specific heat, observed in localized spin systems and known as the Schottky anomaly. In that effect the magnetic energy (or entropy) also shows a non-linear increase vs. temperature.

It can also be mentioned that for $T \to 0$ the thermal expansion coefficient tends to zero, which is a proper thermodynamic behaviour.
On the other hand, in the high temperature region, the anharmonic vibrations play a dominant role and, as a result, an increase in $\alpha_{H,E}$ occurs there in a more regular way.

The next figures (Fig.~\ref{fig9}-\ref{fig14}) are prepared in form of contour graphs, presenting in the multiple charts the deformation (in parts a), magnetostriction coefficient (in parts b) and electrostriction coefficient (in parts c). The isolines of the dimensionless quantities, $\varepsilon$, $1000 \times \lambda_{T,E} t_0$, and $1000 \times \nu_{T,H} t_0/(| e| r_0)$, are plotted as a function of the external fields, in the plane ($H/t_0, \, E | e| r_0/t_0$). Two temperatures, which are far from the ground state: $k_{\rm B}T/t_0=0.5$ (for Fig.~\ref{fig9}, \ref{fig11}, and \ref{fig13}), and $k_{\rm B}T/t_0=1.0$ (for Fig.~\ref{fig10}, \ref{fig12}, and \ref{fig14}) are chosen. In order to illustrate the role of Coulombic on-site repulsion $U$, three values are selected: $U/t_0=2$ for \ref{fig9,fig10}, $U/t_0=5$ for \ref{fig11,fig12}, and $U/t_0=10$ for \ref{fig13,fig14}.

We see that both $T$ and $U$ play a significant role in the behaviour of the quantities presented in multiple contour plots. The shapes of the isolines, defining the areas with similar values of the presented quantities, change dynamically between different plots. In particular, analysing the deformation, it can be noted that areas with the smallest values of $\varepsilon$, denoted by blue shades and situated in the left parts of the charts, where rather small magnetic fields dominate, are shifted quite irregularly along the vertical (electric field) axis for different $T$ and $U$. At the same time, the areas with highest values of $\varepsilon$, denoted by red colour and other warm shades, are expanding as the temperature increases for each value of $U$. 
For instance, analysing the isolines for $\varepsilon$ $=$ 0.1 in Fig.~\ref{fig9}(a)-\ref{fig12}(a) for given $U$, we see that the areas with $\varepsilon \ge$0.1 are increasing when $T$ increases. On the other hand, all isolines in \ref{fig14}(a) correspond to larger values of $\varepsilon$ than isolines in \ref{fig13}(a).
The red areas are situated at high electric fields when $U/t_0=2$ and  $U/t_0=5$, however, when $U/t_0=10$, the area with large $\varepsilon$ extends over all values of the electric fields down to zero, provided the magnetic field is not very small. It should be stressed that deformation $\varepsilon$ is positive for all these cases. The increasing values of $\varepsilon$ when the temperature increases indicate that also the thermal expansion coefficient should be positive.

Regarding the magnetostriction, its value is mostly positive, however, some areas with the negative magnetostriction can be found, namely for high magnetic and electric fields. Interestingly, the area with negative magnetostriction expands with the increasing temperature. On the other hand, the dependency on $U$ is not monotonous: although the area with negative magnetostriction is larger for $U/t_0=5$ than for $U/t_0=2$, however, for $U/t_0=10$ this range becomes strongly reduced and is only visible at higher temperature ($k_{\rm B}T/t_0=1.0$). The area where the highest magnetostriction occurs lies in the range belonging to small electric fields, provided that $U/t_0=2$ or $U/t_0=5$, however, it is shifted towards large electric fields if $U/t_0=10$. The magnetic fields are moderate in all cases of high magnetostriction ranges, being from the range of ($2 \lesssim H/t_0 \lesssim 4$) for $U/t_0=2$, and ($0.5 \lesssim H/t_0 \lesssim 2$) for higher $U/t_0$. The increase in temperature results also in some expansion of the area with strong positive magnetostriction. It can also be noted that isolines corresponding to zero magnetostriction present an interesting behaviour, namely, the increase in electric field causes the decrease in the magnetic field along the isoline, so that the slope of the line as a function of $H$ is negative.

As far as the electrostriction coefficient is concerned, some areas with the negative values of this quantity are also observed in all the cases. For small values of $U$ these areas lie at low electric fields. In particular, for $U/t_0=2$, rather high magnetic fields ($H/t_0 \gtrsim  3$) are involved, whereas for $U/t_0=5$ the corresponding magnetic fields are much lower. They may extend down even until $H/t_0=0$, as in the case of \ref{fig12}. On the other hand, for $U/t_0=10$ the area with negative electrostriction occurs for considerably higher electric fields, for instance, in the order of $E | e| r_0/t_0 \approx 4$, whereas the magnetic fields are still very small. The increase in temperature, from  $k_{\rm B}T/t_0=0.5$ to $k_{\rm B}T/t_0=1.0$, makes the area of negative electrostriction larger, but, at the same time, its amplitude decreases approximately by an order of magnitude. On the other hand, the areas corresponding to the highest positive electrostriction are mainly situated at very large electric fields. An exception is \ref{fig9} where, for $U/t_0=2.0$ and $k_{\rm B}T/t_0=0.5$, the region of positive electrostriction coefficient extends down to very small electric fields. 

In general, the isolines separating the areas of positive and negative electrostriction are more densely packed for $k_{\rm B}T/t_0=0.5$ than for $k_{\rm B}T/t_0=1.0$. The oblique arrangement of these lines at $k_{\rm B}T/t_0=0.5$ may lead to considerable dynamics of the electrostriction coefficient when crossing them horizontally by an increase in the magnetic field. The isolines corresponding to zero electrostriction effect can reveal either positive or negative slope as a function of the magnetic field $H$. The positive slope occurs, first of all, in the range of small magnetic fields, whereas for larger $H$ values the negative slopes are dominating (see, for instance \ref{fig11,fig12} for $U/t_0=5$). This contrasts with the behaviour of the magnetostriction, where the slopes of the null magnetostriction coefficient line were always negative.

 \section{Summary and conclusion}

In the paper the total grand potential has been constructed for the dimer
system, in which the magnetic, elastic and thermal vibration energies play
important roles. The similar attempts at the total Gibbs energy construction,
i.e.,~consisting in summation of various energy contributions, have been
undertaken, so far, for the bulk
materials~\cite{Balcerzak2010,Balcerzak2014a,Balcerzak2017a,Balcerzak2018b,Szalowski2018b}.
According to our knowledge, for the Hubbard dimer such method has not been
presented previously.

It should be mentioned that also another approach to the problem can be
proposed, namely, including the term representing electron--phonon
interaction into the common Hamiltonian, and application of canonical
transformation of Lang--Firsov
type~\cite{Lang_Firsov1962,Stephan1997,Stojanovic2004,Koch2006,Hohenadler2007}.
Then, the procedure is followed by using the 2nd order perturbation calculus, or,
equivalently, by the Schrieffer--Wolff transformation~\cite{Stephan1997}, which
leads to the renormalization of Hamiltonian parameters. Such a method is much
more complicated and, in our opinion, is therefore less convenient for the complete studies
of various thermodynamic properties. Moreover, the Lang--Firsov type transformation
has been proposed for the Hamiltonians describing the coupling of electrons
with the harmonic oscillators~\cite{Stojanovic2004}, whereas in our approach
the anharmonic vibrations are taken into account via Gr\"uneisen parameter
$\Gamma$. It is known that the anharmonicity plays an important role, for
instance, in the studies of thermal expansion effect.

As a result of the present approach, the EOS (Eq.~\ref{eq27}) has been derived for the system in question. This equation yields stable solutions for the interatomic distance deformation, $\varepsilon$, corresponding to the thermodynamic equilibrium state obtained at arbitrary temperature $T$ and the external fields $H$ and $E$. Investigation of this deformation leads to the determination of its derivative quantities such as the thermal expansion or magneto- and electrostriction coefficients. 

\begin{figure}[t]
\begin{center}
\includegraphics[width=0.65\textwidth]{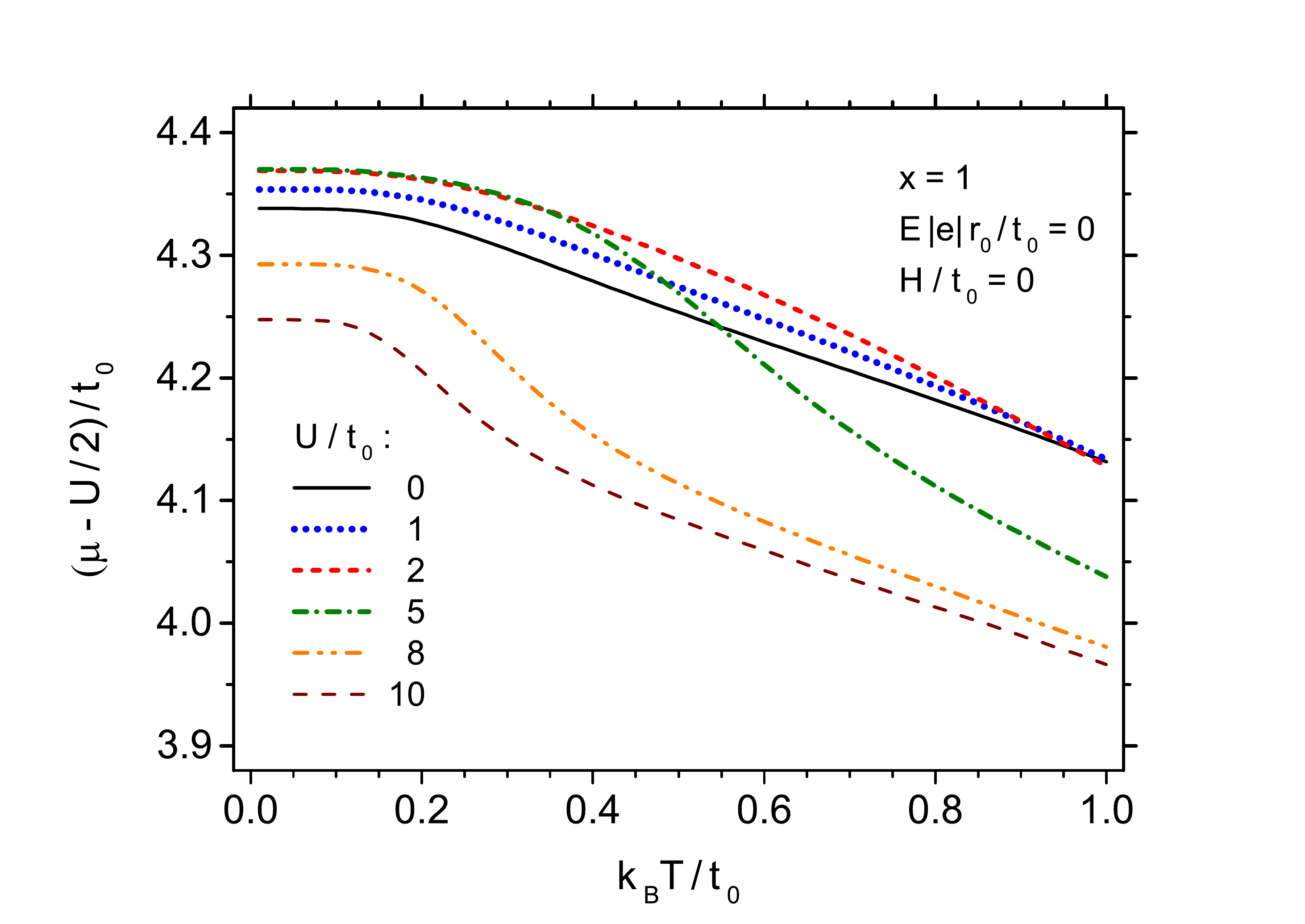}
\caption{\label{fig1}The normalized deviation of the chemical potential from the value of $U/2$ expected for pure Hubbard model as a function of the normalized temperature, for a few selected values of normalized on-site Hubbard energy $U/t_0$ in the absence of the external fields.}
\end{center}

\end{figure}

\begin{figure}[b]
\begin{center}
\includegraphics[width=0.65\textwidth]{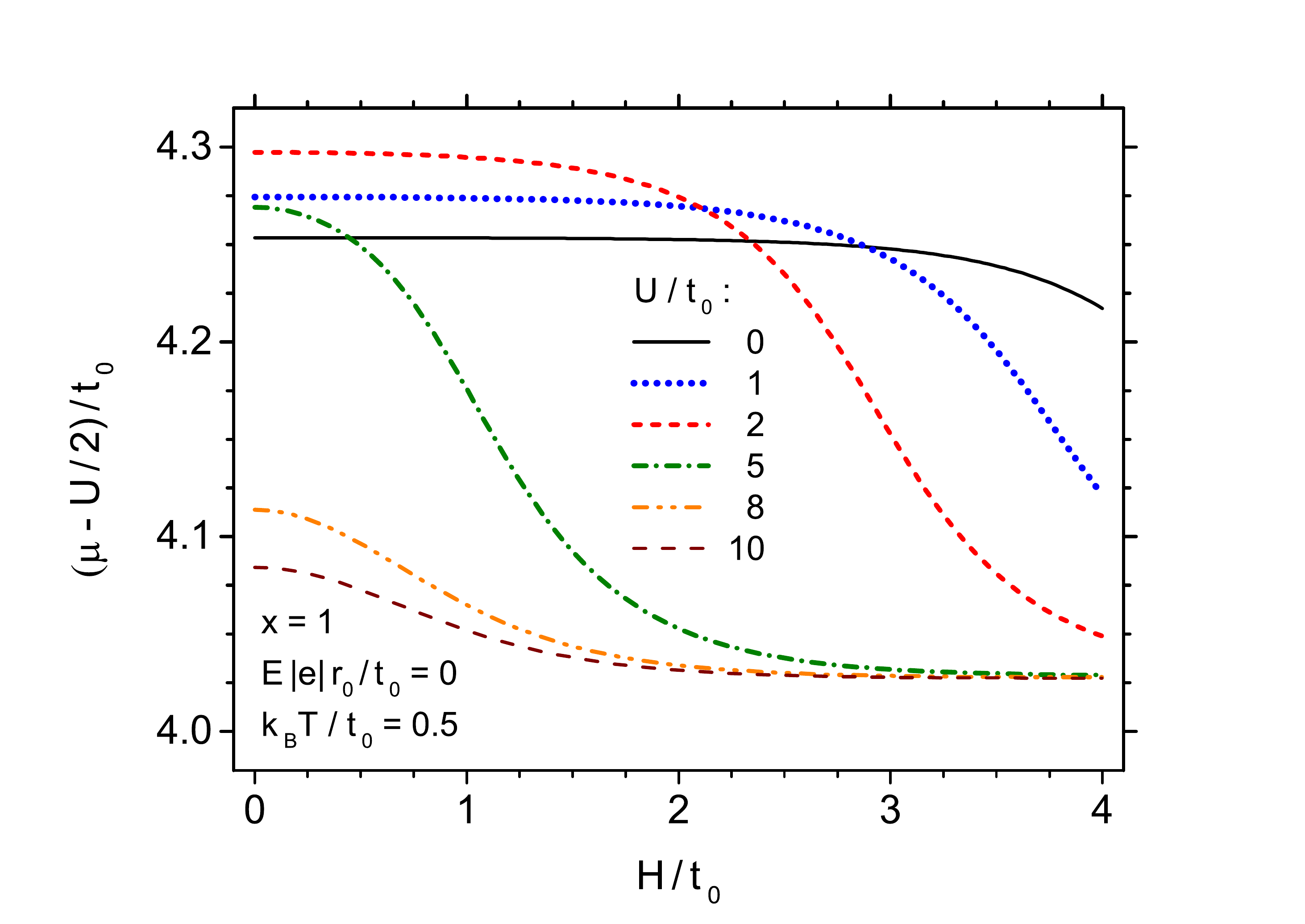}
\caption{\label{fig2}The normalized deviation of the chemical potential from the value of $U/2$ expected for pure Hubbard model as a function of the normalized magnetic field, for a few selected values of normalized on-site Hubbard energy $U/t_0$, in the absence of the electric field and for moderate temperature $k_{\rm B}T/t_0=0.5$.}
\end{center}
\end{figure}

\begin{figure}[t]
\begin{center}
\includegraphics[width=0.65\textwidth]{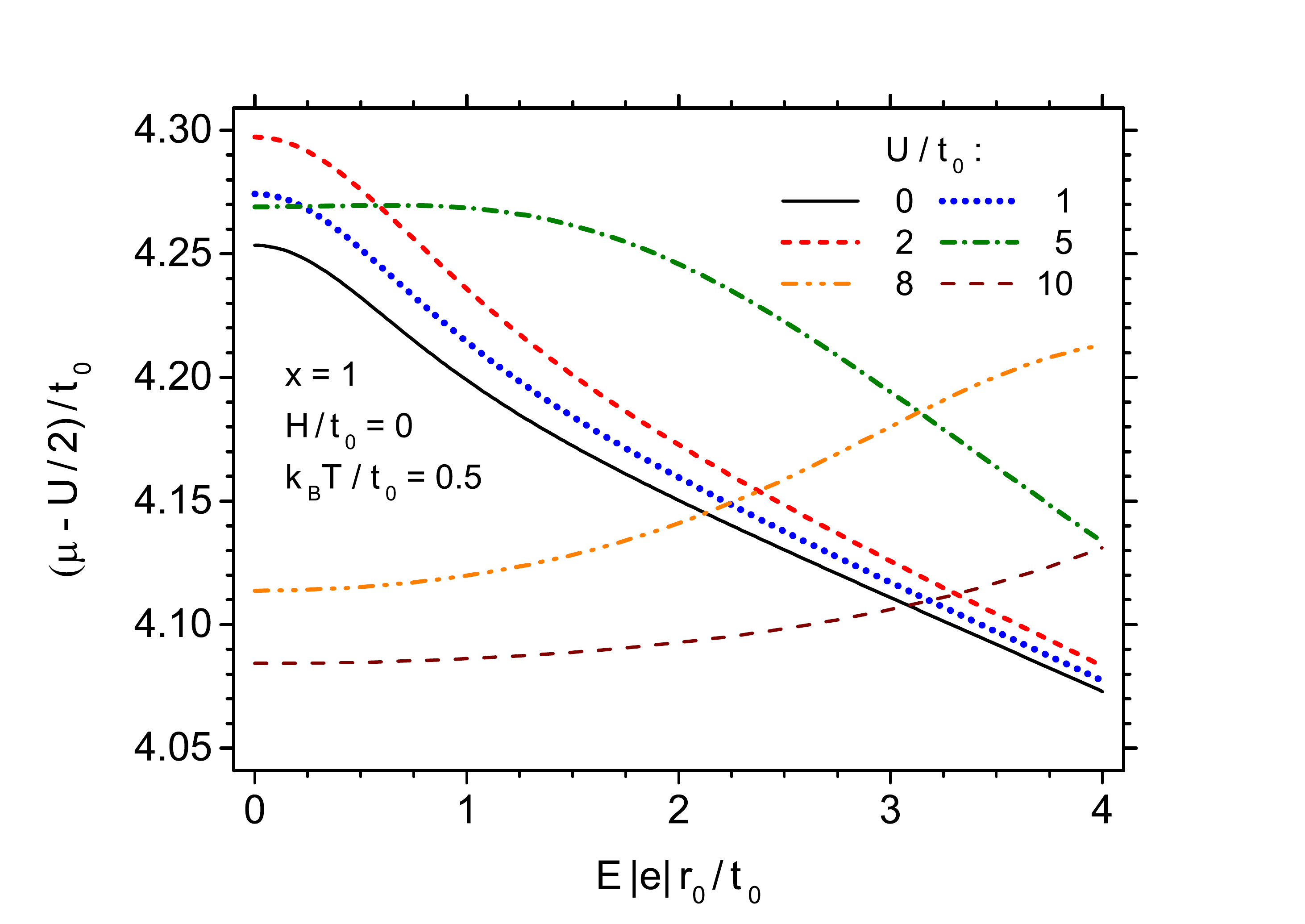}
\caption{\label{fig3}The normalized deviation of the chemical potential from the value of $U/2$ expected for pure Hubbard model as a function of the normalized electric field, for a few selected values of normalized on-site Hubbard energy $U/t_0$, in the absence of the magnetic field and for moderate temperature $k_{\rm B}T/t_0=0.5$.}
\end{center}
\end{figure}

It has been shown in the paper that for the half-filling conditions ($x=1$) and for the elastic interactions included, the chemical potential $\mu$ is not a constant but it depends, via the quantity $\varepsilon$, on the temperature and external fields (Fig.~\ref{fig1}-\ref{fig3}). For $x=1$ the analytic formula describing $\mu$ has been presented (Eq.~\ref{eq21}). 
It should be stressed that the method used here can be equally well applied to arbitrary average concentration of the electrons, i.e.,~when $0\le\; x \le 2$.

\begin{figure}[b]
\begin{center}
\includegraphics[width=0.65\textwidth]{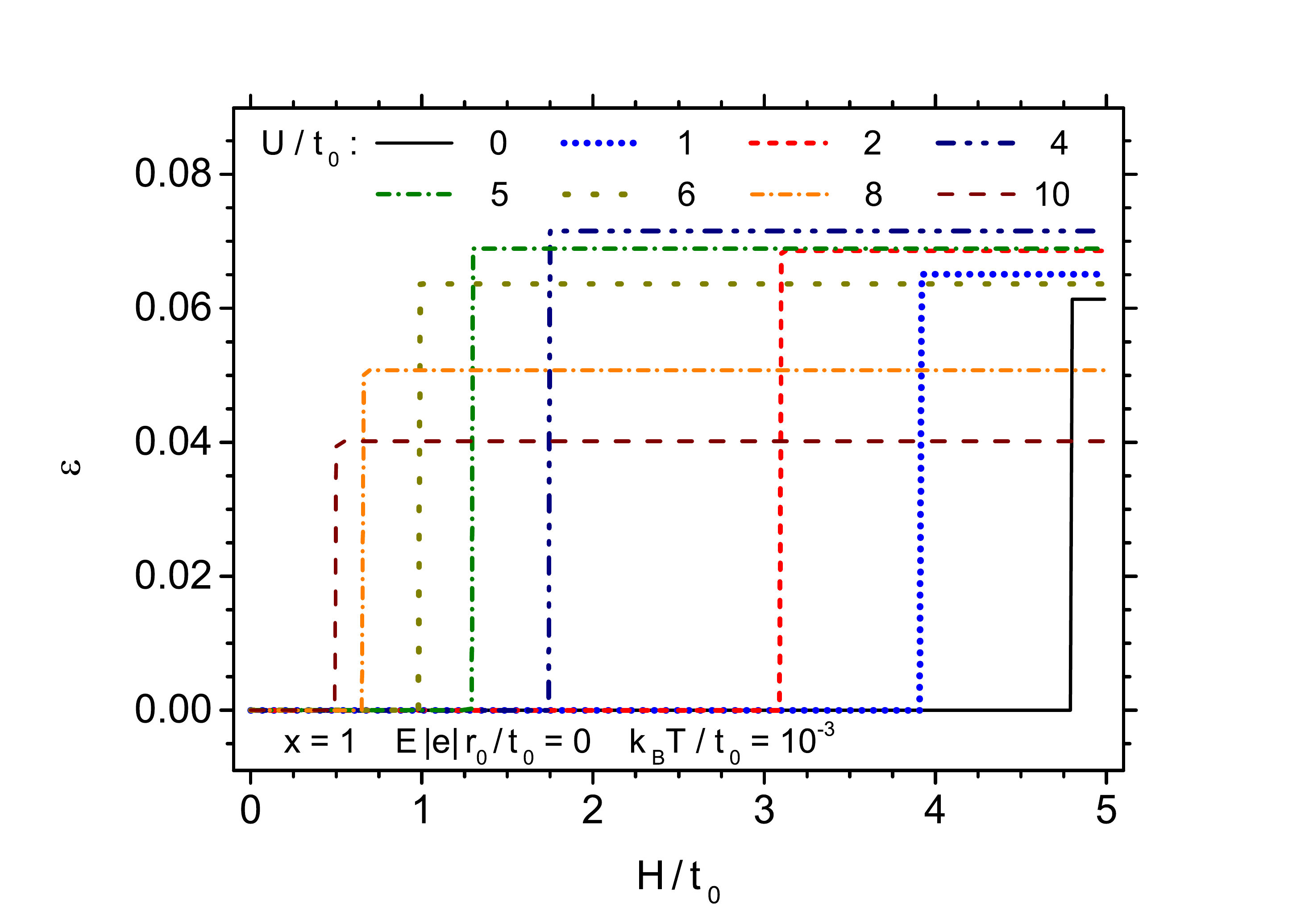}
\caption{\label{fig4}The relative deformation of dimer length as a function of the normalized magnetic field, for a few selected values of normalized on-site Hubbard energy $U/t_0$, in the absence of the electric field and for low temperature $k_{\rm B}T/t_0=10^{-3}$. }
\end{center}
\end{figure}

\begin{figure}[t]
\begin{center}
\includegraphics[width=0.65\textwidth]{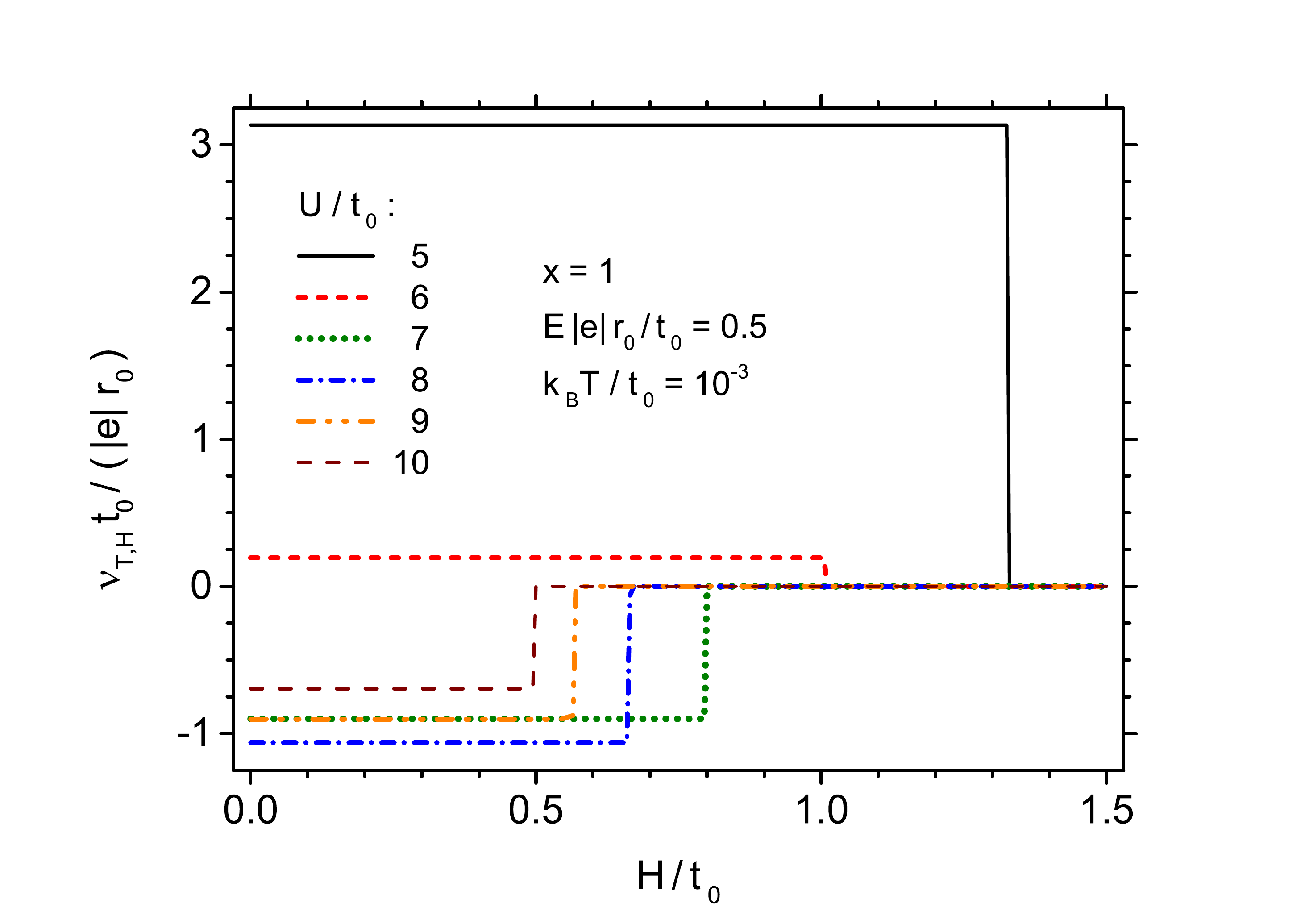}
\caption{\label{fig5}The normalized electrostriction coefficient as a function of the normalized magnetic field, for a few selected values of normalized on-site Hubbard energy $U/t_0$, for the normalized electric field of $E|e|d/t_0=0.5$ and for low temperature $k_{\rm B}T/t_0=10^{-3}$.}
\end{center}
\end{figure}

\begin{figure}[b]
\begin{center}
\includegraphics[width=0.65\textwidth]{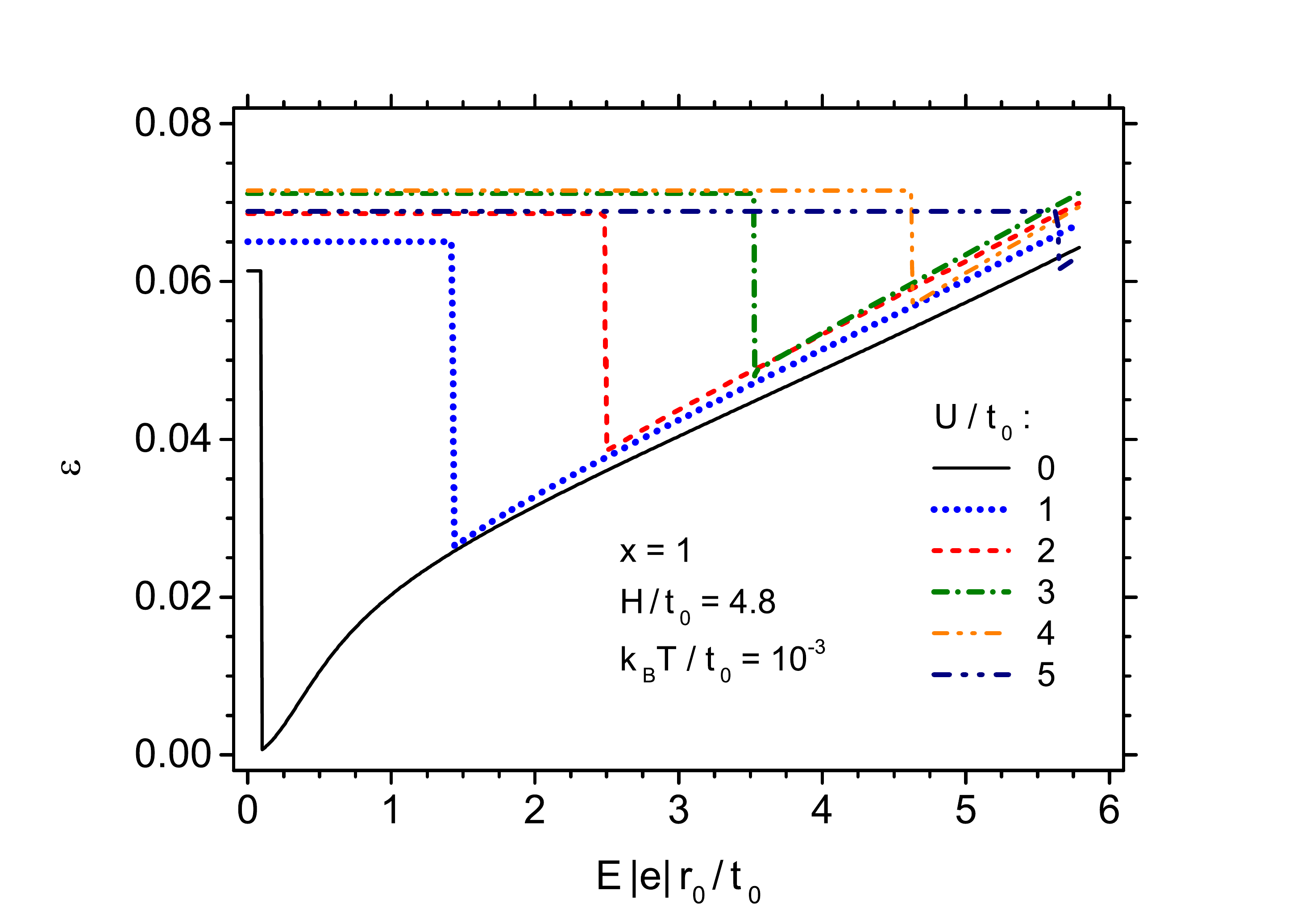}
\caption{\label{fig6}The relative deformation of dimer length as a function of the normalized electric field, for a few selected values of normalized on-site Hubbard energy $U/t_0$, for the normalized magnetic field $H/t_0=4.8$ and for low temperature $k_{\rm B}T/t_0=10^{-3}$.}
\end{center}
\end{figure}

\begin{figure}[t]
\begin{center}
\includegraphics[width=0.65\textwidth]{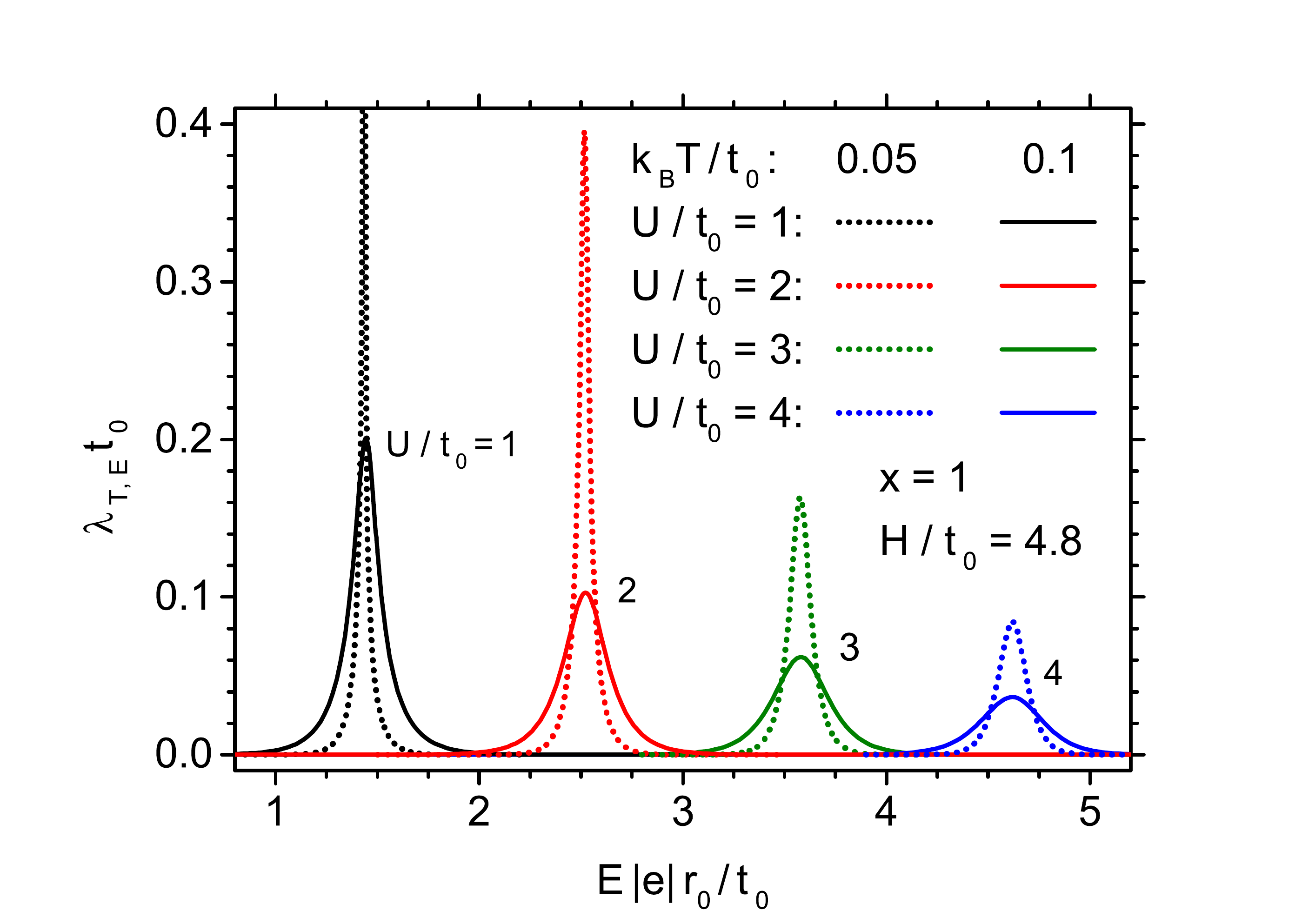}
\caption{\label{fig7}The normalized magnetostriction coefficient as a function of the normalized electric field, for a few selected values of normalized on-site Hubbard energy $U/t_0$, for the normalized magnetic field of $H/t_0=4.8$ and for two normalized temperatures: $k_{\rm B}T/t_0=0.05$ and $0.1$.}
\end{center}
\end{figure}

In the low-temperature region, where the quantum phase transitions are induced by the external fields, the discontinuous changes of deformation $\varepsilon$ have been found at the phase transition points. These discontinuous changes are also accompanied by discontinuities of such quantities as the magnetostriction and electrostriction coefficients (Fig.~\ref{fig4}-\ref{fig7}). It is known that with an increase in temperature the quantum phase transitions become diffused, so then, the most spectacular effects of discontinuity occur at $T \to 0$.

\begin{figure}[b]
\begin{center}
\includegraphics[width=0.65\textwidth]{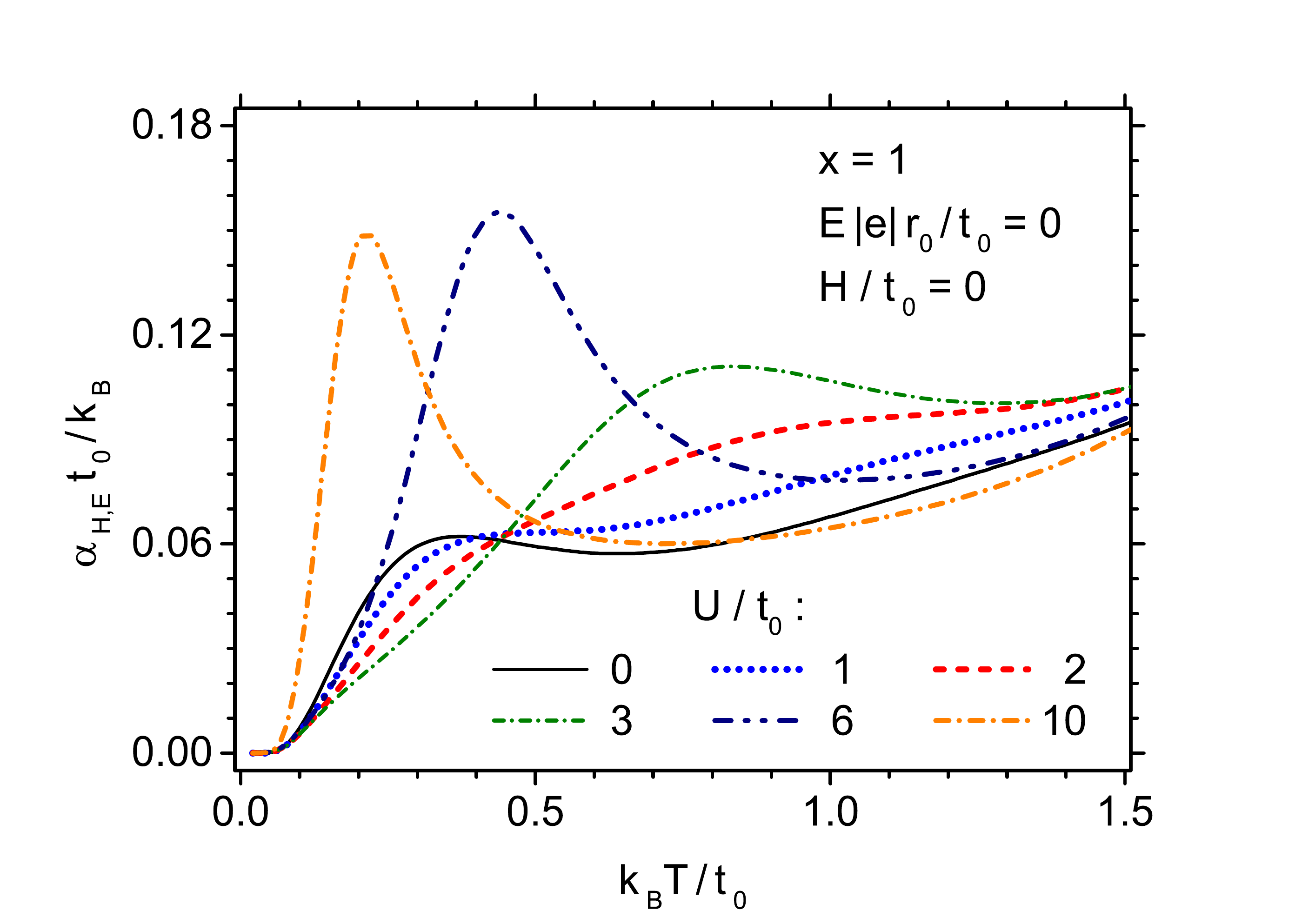}
\caption{\label{fig8}The normalized thermal expansion coefficient as a function of the normalized temperature, for a few selected values of normalized on-site Hubbard energy $U/t_0$, in the absence of the external fields.}
\end{center}
\end{figure}

For the asymmetric interatomic potential, an increase in temperature results in the linear expansivity effect. In our opinion, an interesting finding concerns the broad maxima of the thermal expansion coefficient, found at some temperatures and illustrated in \ref{fig8}. It has been shown that the positions of maxmima are strongly dependent on the values of Hubbard $U$ parameter.

\begin{figure}[h]
\begin{center}
\includegraphics[width=\textwidth]{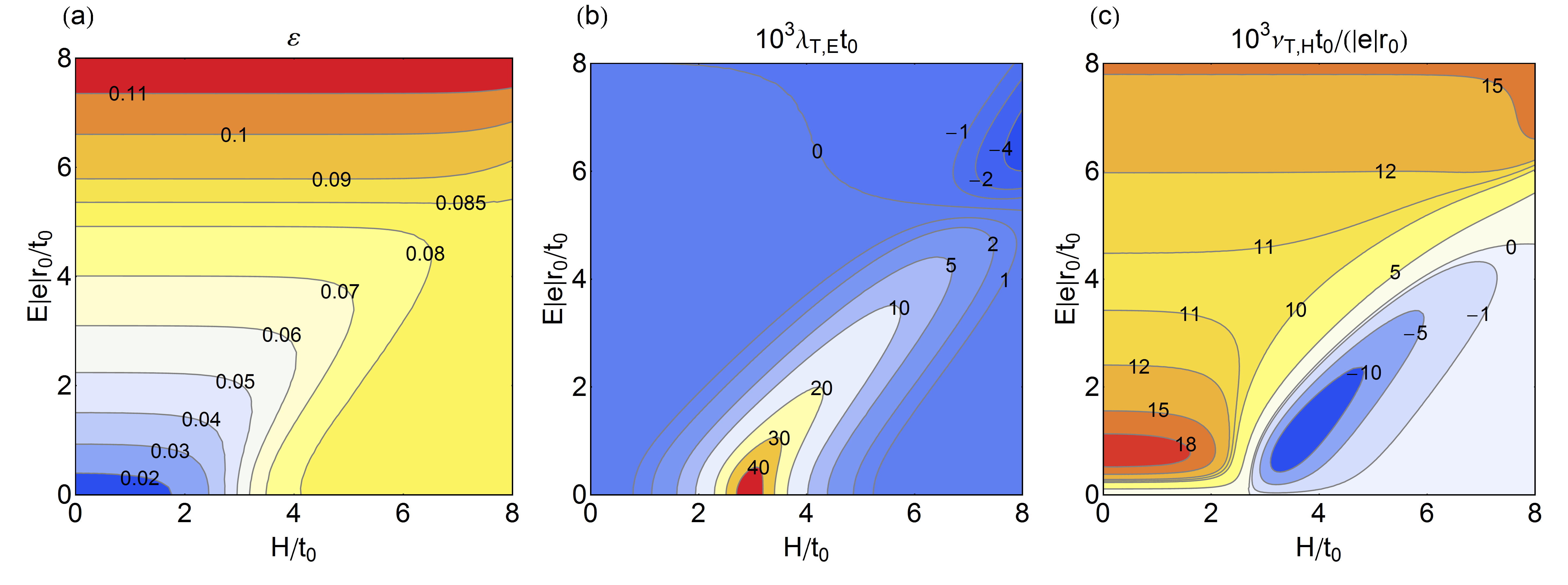}
\caption{\label{fig9}Contour plots of (a) relative deformation of dimer length; (b) normalized magnetostriction coefficient; (c) normalized electrostriction coefficient - as a function of normalized  electric and magnetic field, for $U/t_0=2$ and $k_{\rm B}T/t_0=0.5$.}
\end{center}
\end{figure}

\begin{figure}[h]
\begin{center}
\includegraphics[width=\textwidth]{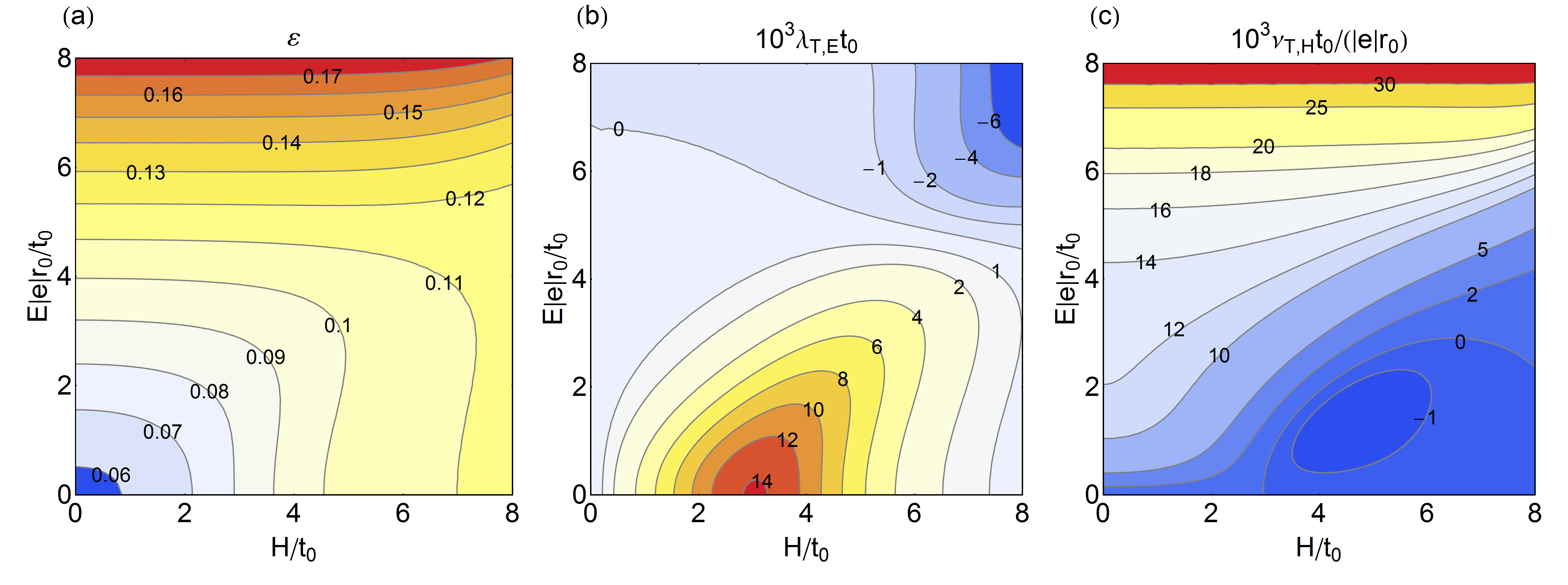}
\caption{\label{fig10}Contour plots of (a) relative deformation of dimer length; (b) normalized magnetostriction coefficient; (c) normalized electrostriction coefficient - as a function of normalized  electric and magnetic field, for $U/t_0=2$ and $k_{\rm B}T/t_0=1.0$.}
\end{center}
\end{figure}

\begin{figure}[h]
\begin{center}
\includegraphics[width=\textwidth]{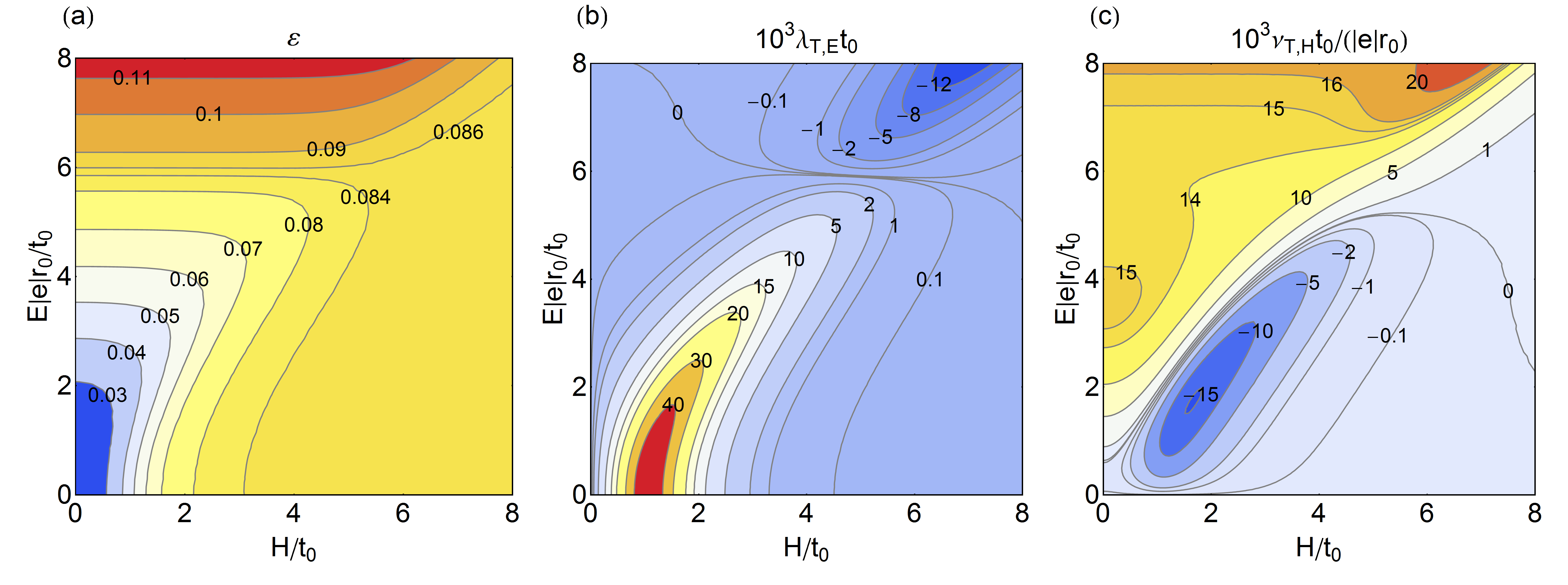}
\caption{\label{fig11}Contour plots of (a) relative deformation of dimer length; (b) normalized magnetostriction coefficient; (c) normalized electrostriction coefficient - as a function of normalized  electric and magnetic field, for $U/t_0=5$ and $k_{\rm B}T/t_0=0.5$.}
\end{center}
\end{figure}

\begin{figure}[h]
\begin{center}
\includegraphics[width=\textwidth]{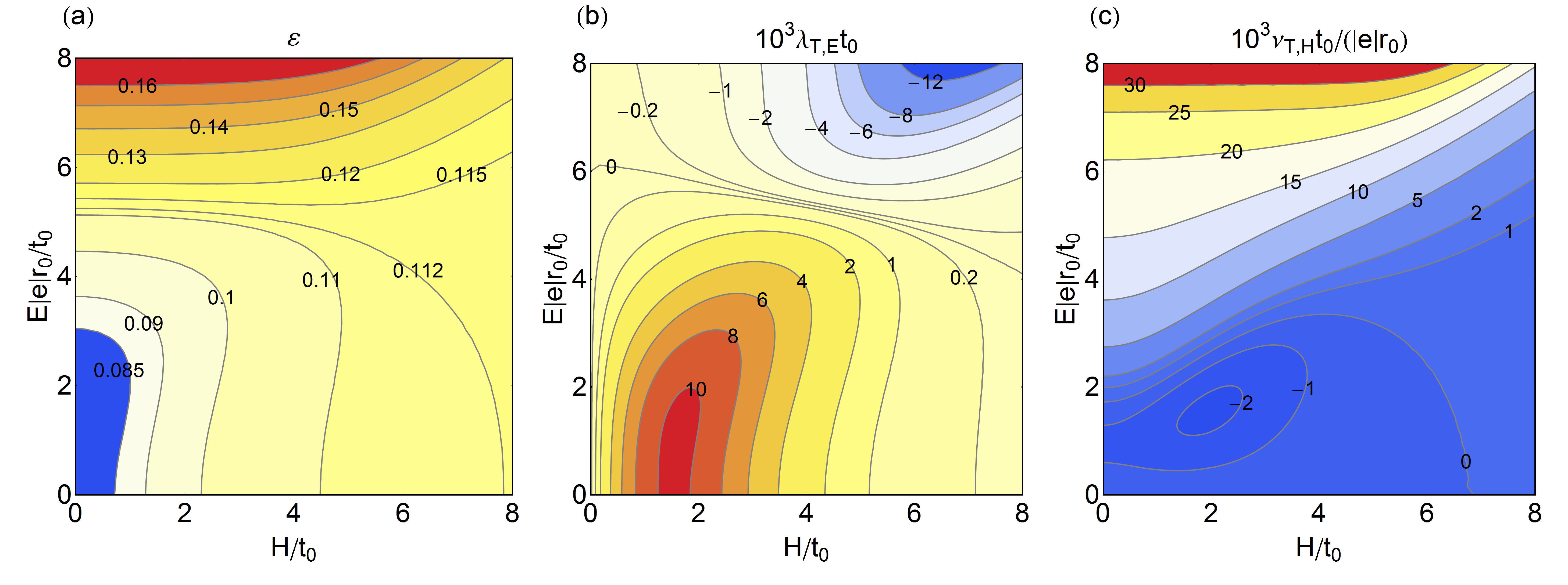}
\caption{\label{fig12}Contour plots of (a) relative deformation of dimer length; (b) normalized magnetostriction coefficient; (c) normalized electrostriction coefficient - as a function of normalized  electric and magnetic field, for $U/t_0=5$ and $k_{\rm B}T/t_0=1.0$.}
\end{center}
\end{figure}

\begin{figure}[h]
\begin{center}
\includegraphics[width=\textwidth]{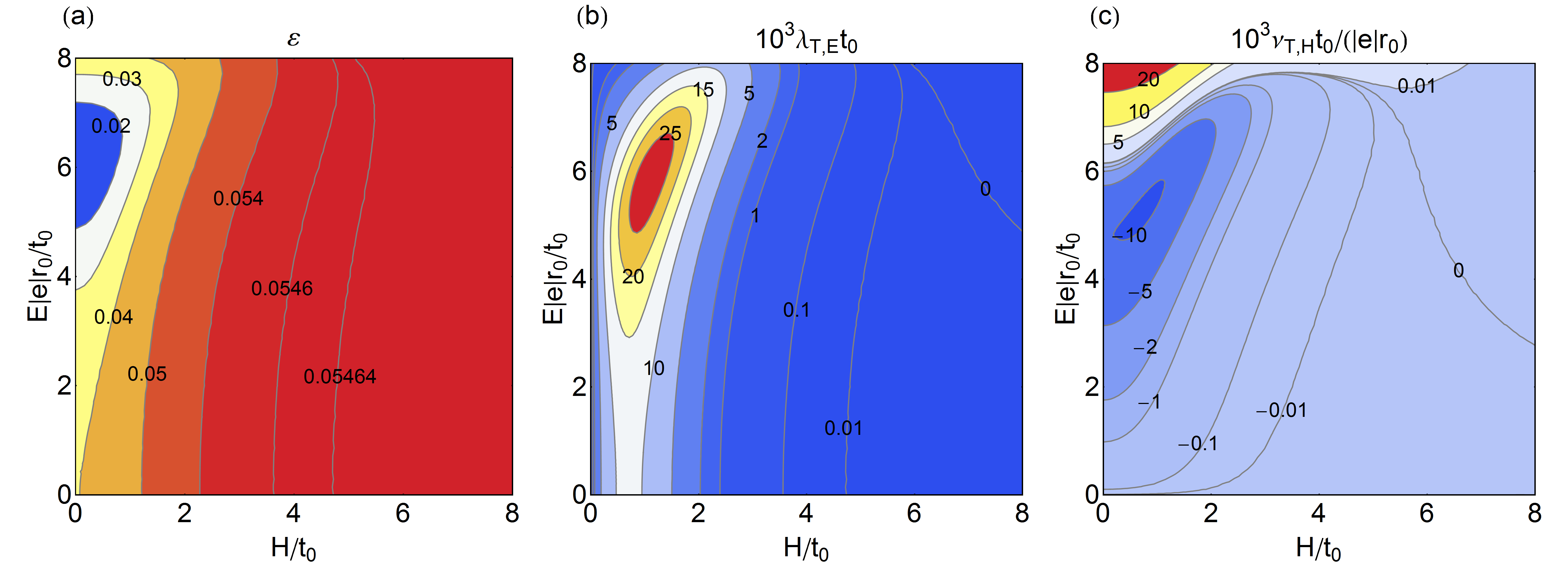}
\caption{\label{fig13}Contour plots of (a) relative deformation of dimer length; (b) normalized magnetostriction coefficient; (c) normalized electrostriction coefficient - as a function of normalized  electric and magnetic field, for $U/t_0=10$ and $k_{\rm B}T/t_0=0.5$.}
\end{center}
\end{figure}

\begin{figure}[h]
\begin{center}
\includegraphics[width=\textwidth]{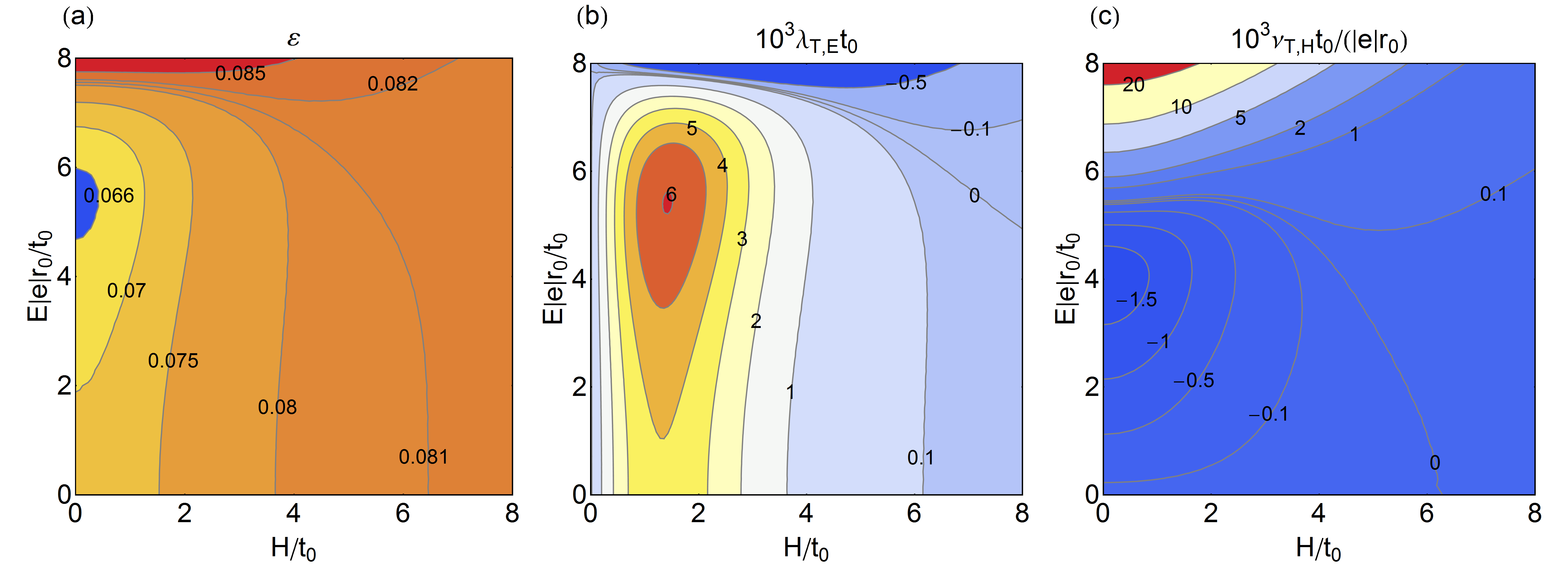}
\caption{\label{fig14}Contour plots of (a) relative deformation of dimer length; (b) normalized magnetostriction coefficient; (c) normalized electrostriction coefficient - as a function of normalized  electric and magnetic field, for $U/t_0=10$ and $k_{\rm B}T/t_0=1.0$.}
\end{center}
\end{figure}

In the high-temperature region, where the discontinuities of thermodynamic quantities do not occur, the contour charts yield a comprehensive description of the interatomic distance deformation and its changes vs. external fields. The contour charts (Fig.~\ref{fig9}-\ref{fig14}) have been used to illustrate simultaneously $\varepsilon$, as well as the magnetostriction and electrostriction coefficients vs. $H$ and $E$, for some selection of temperatures $T$ and parameters $U$. In particular, a distinctive result is an identification of the ($H,E$)-regions where the negative values of these coefficients do appear. It has been found that both temperature $T$ and $U$-parameter have important influence on the localization and area of these regions, as well as on the values of $\varepsilon$, $\lambda_{T,E}$, and $\nu_{T,H}$ themselves.
 
To summarize, by taking into consideration the elastic interactions and thermal
vibrations, a new, more complete thermodynamic description of the Hubbard dimer
has been achieved. It should be stressed that diagonalization of the Hubbard
Hamiltonian has been performed exactly by means of the analytical
methods~\cite{Balcerzak2017}.  It can be expected that selection of other
interatomic potentials would change our results quantitatively; however, the
case which we consider in the present paper captures the essence of the
phenomena emerging due to coupling between purely electronic and elastic
properties.

It should also be pointed out that having constructed the total grand potential, all thermodynamic and statistical properties of the system can be studied. For instance, this allows the self-consistent calculation of the magnetization, electric polarization, magnetic correlation functions, occupation correlations, susceptibility (both magnetic and electric one), entropy and specific heat, in the presence of the elastic interactions.

  

\newpage

\newpage

\newpage

\newpage




\end{document}